\documentclass[aps,pra,preprint,groupedaddress]{revtex4}
\usepackage{color}
\usepackage{epsf}
\usepackage{epsfig}
\usepackage{graphicx}
\usepackage{latexsym}
\usepackage{bm}
\usepackage{amssymb}

\begin{document}


published in: Phys. Rev. A, 88, 042707 (2013)
\title {Ion impact induced Interatomic Coulombic Decay in neon and argon dimers}

\author{H.-K. Kim$^1$}
\email{kim@atom.uni-frankfurt.de}
\author{H.~Gassert$^1$}
\author{M.~S.~Sch\"{o}ffler$^{1}$}
\author{J.~N.~Titze$^1$}
\author{M.~Waitz$^1$}
\author{J.~Voigtsberger$^1$}
\author{F.~Trinter$^1$}
\author{J.~Becht$^1$}
\author{A.~Kalinin$^1$}
\author{N.~Neumann$^1$}
\author{C.~Zhou$^2$}
\author{L.~Ph.~H.~Schmidt$^1$}
\author{O.~Jagutzki$^1$}
\author{A.~Czasch$^1$}
\author{H.~Merabet$^3$}
\author{H.~Schmidt-B\"{o}cking$^1$}
\author{T.~Jahnke$^1$}
\author{A.~Cassimi$^2$}
\author{R.~D\"{o}rner$^1$}

\affiliation{$^1$ Institut f\"ur Kernphysik, University Frankfurt,Max-von-Laue-Str. 1, 60438 Frankfurt, Germany}
\affiliation{$^2$ CIMAP Caen, GANIL, Bd Henri Becquerel, BP 55027 – 14076 Caen Cedex 05, France}
\affiliation{$^3$ Department of Mathematics, Statistics and Physics, College of Arts and Sciences, P.O. Box: 2713 Doha, Qatar}

\vskip 5mm

\date{\today}

\vskip 5mm

\begin{abstract}
We investigate the contribution of Interatomic Coulombic Decay induced by ion impact in neon and argon dimers (Ne$_{2}$ and Ar$_{2}$) to the production of low energy electrons. 
Our experiments cover a broad range of perturbation strengths and reaction channels. We use 11.37 MeV/u S$^{14+}$, 0.125 MeV/u He$^{1+}$, 0.1625 MeV/u He$^{1+}$  and 0.150 MeV/u He$^{2+}$ as projectiles and study ionization, single and double electron transfer to the projectile as well as projectile electron loss processes. The application of a COLTRIMS reaction microscope enables us to retrieve the three-dimensional momentum vectors of the ion pairs of the fragmenting dimer into Ne$^{q+}$/Ne$^{1+}$ and Ar$^{q+}$/Ar$^{1+}$ (q = 1, 2, 3) in coincidence with at least one emitted electron.
\end{abstract}


\maketitle


\section{Introduction \label{introduction}}
In ionizing ion atom collisions the emitted electrons typically show a continuous energy distribution with decreasing intensity towards higher energies. The dominant contribution to the electron energy distribution results mainly from distant collisions with small energy transfers which can be approximated by an exponential decay \cite{Rudd1976,Rudd1979}.
Recently we showed for He$^{+}$/Ne$_{2}$ collisions \cite{kim2011} that for Van der Waals dimers there is a significant increase of low energy electrons on top of the distribution for atomic targets. 

Such low energy electrons have been shown to efficiently cause double strand breaks in DNA \cite{boudaiffa2000,hanel2003,martin2004,sanche2005}. We could trace this enhanced production of low energy electrons back to an autoionization mechanism termed Interatomic Coulombic Decay (ICD). In ICD an excited target atom (e.g. by the removal of a 2s-electron in neon) relaxes into the ground state by transferring its excess energy to another neutral atom where a low energy (ICD) electron is emitted. In contrast to isolated excited atoms in which radiative decay occurs on a long time scale the ICD channel is open when the excited atom is located in close vicinity to other atoms. ICD has therefore been suggested to be a relevant mechanism in causing radiation damage to healthy tissue. In return ICD was also proposed to be utilized in radiation therapy by specifically triggering it in the proximity of tumor cells via resonant Auger decay \cite{gokhberg2013}.

ICD was predicted in 1997 by Cederbaum et al. \cite{cederbaum1997} and experimentally confirmed in 2003 by Marburger et al. \cite{marburger2003} for neon clusters and in 2004 for neon dimers by Jahnke et al. \cite{jahnke2004}. Today "it is now well-known that ICD appears everywhere and transfers the energy and the charge from the species with the vacancy to the environment surrounding it." \cite{ouchi2011}. This statement refers to the divers experimental and theoretical studies on ICD occurring in many systems. It has been observed or predicted for extremely large quantum objects as the helium dimer \cite{havermeier2010,sisourat2010}, biologically relevant aqueous systems such as water clusters \cite{jahnke2010,mucke2010} and liquid water \cite{aziz2008}, quantum dots \cite{bande2011}, surfaces \cite{grieves2011} and a vast number of rare gas clusters \cite{jahnke2007a,lablanquie2007,ueda2007,morishita2008,stoychev2008,ueda2008,kreidi2008a,kreidi2008b,dias2008,demekhin2008,kreidi2009,arion2010,havermeier2010b,ouchi2011b,sakai2011,semenov2012}   (See recent reviews and references therein \cite{averbukh2011,hergenhahn2011}). These works differ in the chosen "target system" but also in the way the excitation is created experimentally. The most common route to create the excitation is by an initial valence or inner-shell photoionization by synchrotron radiation but also resonant excitation or, as was shown very recently, resonant Auger excitation of the target \cite{kimura2013,trinter} has proven to initiate ICD. There are also two very recent studies in which ICD was observed after ion impact \cite{kim2011,titze2011}. 
In the present work we give a comprehensive overview of the observation of ICD (and its significant contribution to the low energy electron yield) in ion/dimer collisions for He$^{1+}$ and He$^{2+}$ projectiles at energies from the maximum of the stopping power of liquid water (v $\approx$ 2.5 a.u.) \cite{garciamolina2009,palmer1978,Haque1985} up to very fast (v = 21.2 a.u. and q/v = 0.66) S$^{14+}$ projectiles where the momentum transfer between the projectile and the target system is small \cite{ullrich2003}. We observe multiple breakup channels for collisions with and without change of the projectile charge state in neon and argon dimers. The change of the projectile charge state $\Delta$q$_{proj}$ = q$_{proj,i}$ - q$_{proj,f}$ is one way to categorize the reaction, with q$_{proj,i}$ being the initial and q$_{proj,f}$ the final projectile charge state. The processes are referred to as ionization I ($\Delta$q$_{proj}$ = 0), projectile loss PL ( $\Delta$q$_{proj}$ = -1), single capture SC ($\Delta$q$_{proj}$ = 1) and double capture DC ($\Delta$q$_{proj}$ = 2).

\section{Experiment \label{experiment}}

We used a COLTRIMS \footnote{cold target recoil ion momentum spectroscopy} \cite{doerner2000,ullrich2003} reaction microscope to measure the momenta of the fragments (see table \ref{configuration} for experimental parameters). The projectile ion beam and the target gas jet are intersected at an angle of 90 degrees forming an interaction region of a few mm$^{3}$. The electrons and ions created in the collision are guided by a weak homogeneous magnetic and electric field onto time- and position-sensitive channel-plate detectors with delay-line readout (PSD) \cite{jagutzki2002a,jagutzki2002b} \footnote{80 to 120 mm active diameter, depending on the collision system under study}. The particles are collected with a 4 $\pi$ solid angle for the displayed energy ranges. Due to an electron detection efficiency of less than 0.4 the number of detected events decreases with a scaling factor of at least 0.4$^{N}$ (N: number of electrons). In the data analysis we therefore consider both ions and one of the emitted electrons. For the ionization channel without charge exchange the projectile could not be detected and a pulsed ion beam was used to measure the times-of-flight. For those channels where the projectile changed its charge state a continuous ion beam was used and the projectile was charge state analyzed by an electrostatic deflector and detected on a third PSD. This delivered a stop signal for the time-of-flight measurement. The three-dimensional momentum vectors and therewith the kinetic energies of the particles are reconstructed from the particles' times-of-flight and positions of impact on the detector. The precooled (T$_{0}$ = 150 - 300 K) target gas is further cooled in a supersonic gas expansion through a nozzle (d$_{nozzle}$ = 5 - 30 $\mu$m) at a driving pressure p$_{0}$ of a few bar. Two skimmers (\textless  0.5 mm) reduce the gas jet diameter to 1 - 1.5 mm at the interaction region. Depending on the particular experimental parameters a dimer to monomer fraction of approximately 1\% is estimated. The projectiles with an energy of E$_{proj,S14+}$ = 11.37 MeV/u and a relative energy spread of approximately 10$^{-4}$ were delivered by the heavy ion accelerator facility "GANIL" (Caen, France) at the intermediate energy beamline "SME" in cave D at intervals of 81.2 ns. The projectiles were provided in bunches of 2 mm in diameter. The bunch length at the target position is about 1 ns. This is not deteriorated by the energy spread of the beam, since there is an active time focusing unit in the beamline which compensates for this spread in time due to the velocity spread.
 The electron times-of-flight are typically in the order of a few ten nanoseconds (20 - 50 ns) and with that smaller than the bunch period. The ion times-of-flight (1 - 10 $\mu$s) are thus unambiguously determined in an electron-ion-bunchmarker coincidence.
The helium projectile ions at "low" energy were generated by the Van de Graaff accelerator at the Institut f\"ur Kernphysik of the Goethe-Universit\"at in Frankfurt as described in \cite{kim2011} for the ionization experiment and as described in \cite{titze2011,kim2011} for the experiments with projectile charge state change. The typical recoil momenta of the ions in such collisions are below 2 a.u. \cite{doerner2000} and thus can be neglected compared to the momenta gained in the Coulomb explosion (up to 200 a.u.).

\begin{table}
\begin{center}
\begin{tabular}{cccccccccc}
\hline
\ Configuration &  Energy & Projectile & Target & E-field & B-field &  T$_{0}$ & p$_{0}$ & d \\
\ & [MeV/u] &  & & [V/cm] & [G] &   [K] &  [bar]  & [$\mu$m]\\
\hline
I)& 0.125& He$^{1+}$&Ne$_{2}$&29.02&14.06&150&6&30\\
II)& 0.1625& He$^{1+}$&Ne$_{2}$& 6.14&7.65&180&6&30\\
III)& 0.150& He$^{2+}$ &Ne$_{2}$&22.6&13.85&150&10&30\\
IV)& 0.150& He$^{2+}$ &Ar$_{2}$&22.6&13.85&300&1.5&30\\
V)& 11.37 & S$^{14+}$&Ne$_{2}$&24.49&15.2&150&11&5\\
\hline
\end{tabular}
\caption[configuration]{The configuration for the extraction fields and the parameters for the target gas jet are given for each projectile and target gas whereby T$_{0}$ is the temperature of the precooled gas, p$_{0}$ the driving pressure and d the diameter of the nozzle.}
\label{configuration}
\end{center}
\end{table}

\section{Results and Discussion \label{results}}

\vskip 24pt

\subsection*{Experimental identification of Interatomic Coulombic Decay}
In this section we discuss how events from ICD can be distinguished from other ionization pathways in our experiments. Interatomic Coulombic Decay is an autoionization process by which an excited atom decays transferring its excess energy to a neighboring atom. A clear illustration for ICD is given by Jahnke et al. \cite{jahnke2004} for the neon dimer which is composed of two neon atoms bound by the Van der Waals force at an equilibrium internuclear distance of R$_{0,Ne2}$ = 3.1 \AA\ \cite{wuest2003} with a binding energy of \textless E$_{bind,Ne2}$\textgreater = 3 meV. 
Here an excited state of the neon dimer is created by the removal of an innervalence 2s electron (ionization potential IP$_{2s}$= 48.47 eV). This leaves one constituent in the cationic Ne$^{+}$(2s$^{1}$2p$^{6}$) state and the other in the neutral ground state Ne(2s$^{2}$2p$^{6}$). An outervalence electron of the cation fills the 2s-hole transferring the released energy (26.91 eV) to the neutral atom where a 2p outervalence electron (IP$_{2p}$ = 21.56 eV), the low energy ICD electron (e$_{ICD}$), is ejected. Both ions are then singly positively charged and are driven apart by their Coulomb repulsion gaining a final kinetic energy (E$_{KER}$: Kinetic Energy Release) which is observed. In this particular system ICD is unambiguously identifiable in a coincident electron-ion-ion measurement as the sum of the particles’ kinetic energies is a constant of E$_{sum}$ = 5.35 eV = 48.47 eV - 2 $\cdot$ 21.56 eV.
This is the total energy that is released when the bound dimer state Ne$^{1+}$(2s$^{-1}$)[$^{2}$S$_{1/2}$] + Ne[$^{1}$S$_{0}$] decays to the dissociative Ne$^{1+}$(2p$^{-1}$)[$^{2}$P$_{1/2,3/2}$] + Ne$^{1+}$(2p$^{-1}$)[$^{2}$P$_{1/2,3/2}$] state. In Fig. 1 the E$_{KER}$ and the energy of the electron is shown for an ion impact experiment with electron transfer to the projectile (single capture) with
 \begin{equation}\label{E1}
(0.1625 MeV/u)He^{1+}+Ne_{2}\rightarrow He^{0}+Ne^{1+}(2p^{-1})+Ne^{1+}(2p^{-1})+e^{-}
 \end{equation}
being the reaction equation. 
\begin{figure*}[ht]
\centerline{\psfig{figure=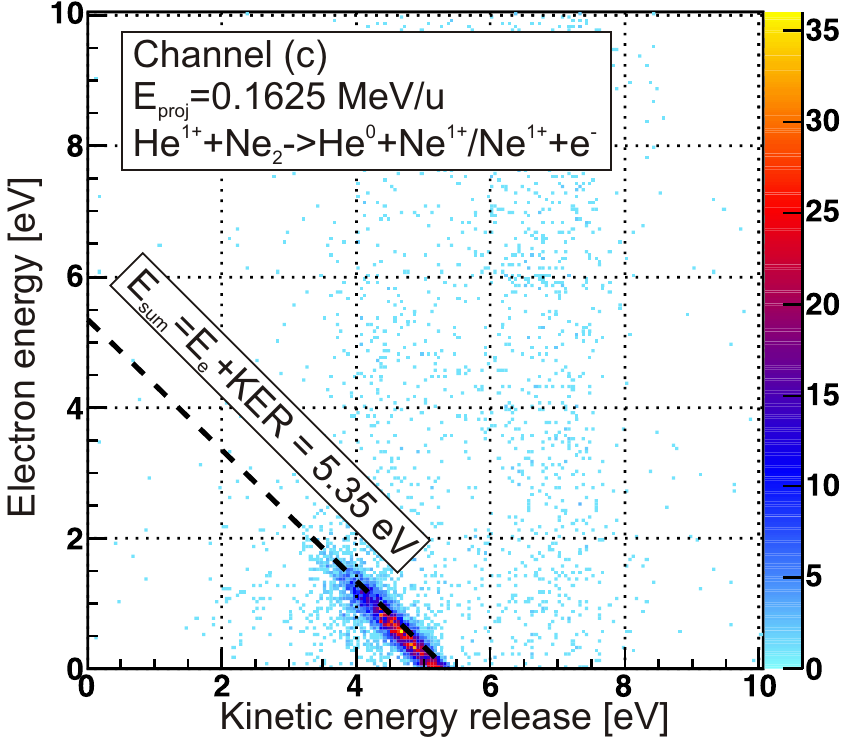,width=0.5\textwidth}}
\caption{(Color online) Correlation of electron and ion energies for reaction channel eq. \ref{E1} (later referred to as channel c)). Horizontal axis: sum energy of both recoil ions (kinetic energy release), vertical axis: energy of the electron. Events from ICD are distributed along a diagonal line (see text).} \label{channelc}
\end{figure*}

The dominant diagonal structure at E$_{sum}$ = 5.35 eV, also seen in \cite{jahnke2004}, can be attributed to the events in which the electron is released via ICD (the other electron has been captured by the projectile). Apart from this diagonal line the histogram also shows additional events that are spread continuously over the displayed electron energy range and which are located at 3.5 eV \textless\ E$_{KER}$ \textless\ 8 eV with a minimum for E$_{KER}$ $\approx$ 5.5 eV. These events can be attributed to two additional reaction mechanisms, which were reported to occur in argon dimers \cite{matsumoto2010,matsumoto2011}, helium dimers \cite{titze2011} and in neon dimers \cite{kim2011} after ion impact. In the first process the dimers' constituents are ionized by two consecutive interactions with the projectile (interaction in two steps) leading to Coulomb fragmentation. For fast projectiles these collisions take place on a subfemtosecond timescale which is much shorter than the timescale for nuclear motion. Thus this direct Coulomb explosion (CE) occurs almost at the equilibrium internuclear distance of the ground state dimer. According to the reflection approximation \cite{gislason1973} the KER for the ground state bond length of the neon dimer R$_{0,Ne2}$ = 3.1 \AA\ is given by  E$_{KER}$ = q$_{1}$q$_{2}$/R$_{0,Ne}$ = 4.6 eV, thus CE contributes to the events located at 3.5 eV \textless\ E$_{KER}$ \textless\ 5.5 eV. As discussed above this is about the same region in E$_{KER}$ populated by ICD. This degeneracy in the E$_{KER}$ between CE and ICD is present for all ICD channels with short life-times where the Coulomb repulsion also takes place without significant preceding nuclear motion. For this particular ICD channel in neon the life-time was predicted to be between 85 fs \cite{santra2001} and 168 fs \cite{vaval2007}. The second process creating two charged fragments is referred to as the radiative charge transfer (RCT). Here only one site of the dimer is doubly ionized (Ar$^{2+}$/Ar and Ne$^{2+}$/Ne) in an initial interaction with the projectile (interaction in one step). This doubly charged dimer with the charges both on one site then contracts and when the shells of both atoms start to overlap charge transfer between the two sites becomes possible. The excess energy is emitted by a photon. The subsequently formed ion pair (Ar$^{1+}$/Ar$^{1+}$ and Ne$^{1+}$/Ne$^{1+}$) then also Coulomb explodes. Due to the preceding contraction of the dimer these events are located at higher kinetic energy releases (In Fig. \ref{channelc} for neon dimers at 5.5 eV \textless\ E$_{KER}$ \textless\ 8 eV). 

\begin{figure*}[ht]
\centerline{\psfig{figure=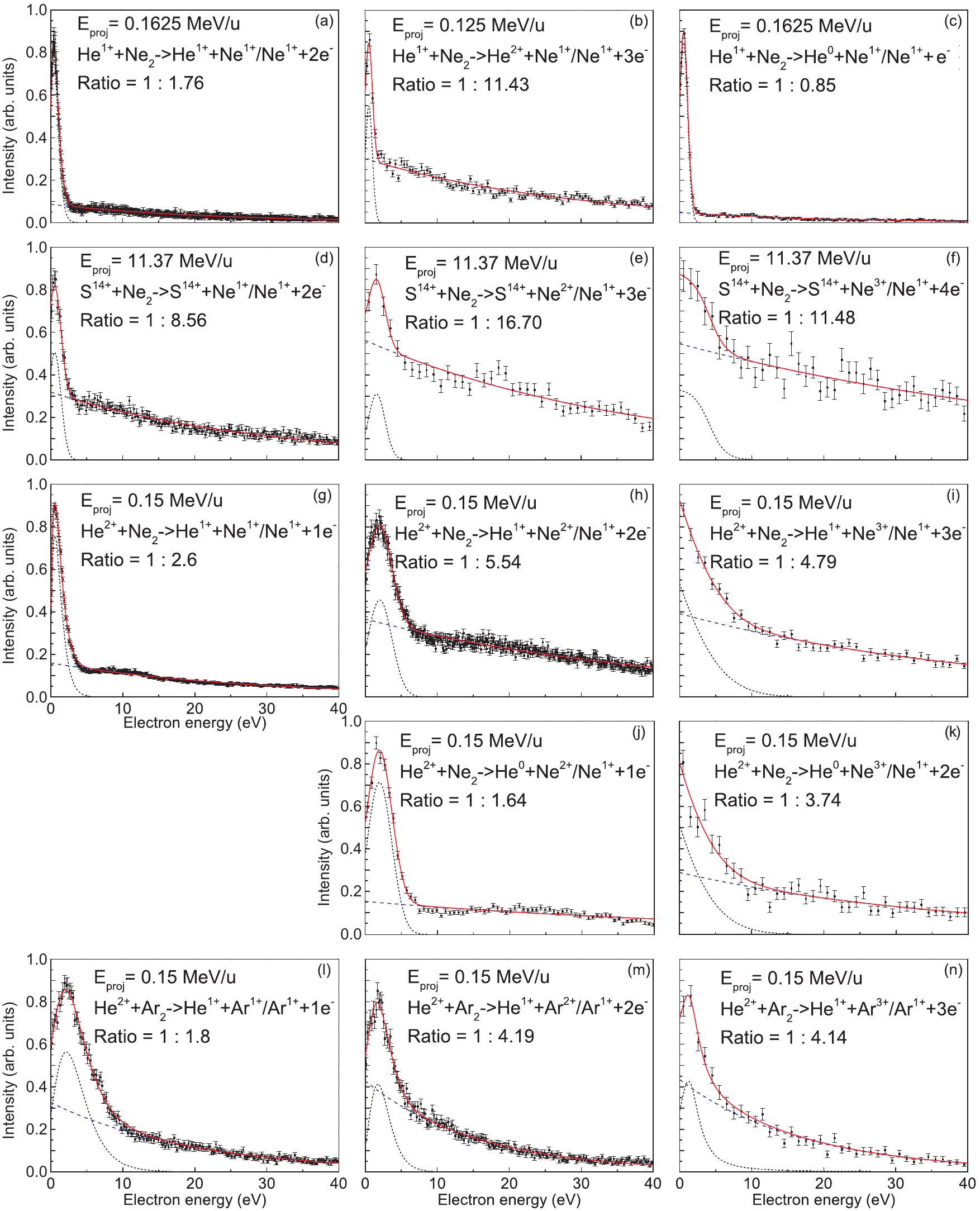,width=1.0\textwidth}}
\caption{(Color online) Electron energy distribution created in the fragmentation channels listed in table \ref{channels}. The fraction for the ICD electrons (e$_{ICD}$) to the continuous electrons (e$_{cont}$) is given by the ratio N$_{e,ICD}$/N$_{e,cont}$. N$_{e,cont}$ and N$_{e,ICD}$ were extracted by integrating the areas of two functions fitted to the peak (black short-dashed line) and to the exponentially decaying continuous distribution (blue long-dashed line) for the displayed electron energy range. The red solid line is the sum of the black and blue curve.} \label{ee}
\end{figure*}

If one integrates over all E$_{KER}$ one obtains electron energy distributions as shown in Fig. \ref{ee}a - n (A detailed description is given in section "reaction channels".). These electron spectra show the contribution of ICD to the total production of electrons in the collision. The narrow peak at low energies in Fig. \ref{ee}c corresponds to the diagonal line in Fig. \ref{channelc} and hence is created by ICD while the other channels produce a broad exponentially decaying energy spectrum well-known for ion and electron impact ionization. To estimate more quantitatively the fraction of ICD electrons we have fitted the two spectral features and find a ratio of ICD electrons to all other electrons of N$_{e,ICD}$ to N$_{e,cont}$. All collision systems which we studied show electron energy distributions with similar shapes (as depicted in Fig. \ref{ee}) which can be qualitatively decomposed into two parts. 
The first part are the electrons which are created in an initial interaction with the projectile and which are spread continuously over the whole electron energy range. Other than in photoionization experiments with a discrete photoelectron energy this behaviour is typical for ion induced ionization due to the continuous momentum and energy transfer. The second part is an additional peak structure at low energies on top of the continuous distribution which results from the secondary ICD electrons. A rough estimation of the fraction for the ICD electrons (e$_{ICD}$) to the continuous electrons (e$_{cont}$) is given by the ratio N$_{e,ICD}$/N$_{e,cont}$. N$_{e,cont}$ and N$_{e,ICD}$ were extracted by integrating the areas of two functions fitted to the peak (black short-dashed line) and to the exponentially decaying continuous distribution (blue long-dashed line) for the displayed electron energy range.

\subsection*{Reaction channels}
An overview of the reaction channels studied is given in table \ref{channels}. Here the projectile energy, initial and final projectile charge state, the final charge state of the recoil ion pair, the ratio of electrons originating from ICD to other electrons, and the experimental configuration is given. In channels a) - c) He$^{1+}$ projectiles at energies from 0.125 MeV/u - 0.1625 MeV/u were utilized to induce Ne$^{1+}$/Ne$^{1+}$ breakups by different mechanisms of electron removal. These include at least one ionization, single capture and projectile loss processes. In d) - f) S$^{14+}$ projectiles at an energy of 11.37 MeV/u were used to ionize the neon dimer leading to Ne$^{1+}$/Ne$^{1+}$, Ne$^{2+}$/Ne$^{1+}$ and Ne$^{3+}$/Ne$^{1+}$ ion pairs. The same final ion charge states were measured for neon dimer fragmentation with single (g) - i)) and double electron capture (j) - k)) events into (0.15 MeV/u) He$^{2+}$ projectiles \footnote{The Ne$^{1+}$/Ne$^{1+}$ channel for double capture is not discussed, since no ICD electron can be emitted.}. Finally in l) - n) analogous processes to g) - i) were observed in argon dimers.

\begin{table}
\begin{center}
\begin{tabular}{ccccccc}
\hline
\ Channel &  Energy & Projectile$_{i}$ & Projectile$_{f}$& Recoil ion pair & Ratio& Configuration \\
\ &  [MeV/u] &  & & & & \\
\hline
a)\footnote{same data as in \cite{kim2011}}& 0.1625& He$^{1+}$& He$^{1+}$& Ne$^{1+}$/Ne$^{1+}$&0.57&II)\\
b)& 0.125& He$^{1+}$& He$^{2+}$& Ne$^{1+}$/Ne$^{1+}$&0.09&I)\\
c)& 0.1625& He$^{1+}$& He$^{0}$& Ne$^{1+}$/Ne$^{1+}$&1.18&II)\\
d)& 11.37& S$^{14+}$& S$^{14+}$& Ne$^{1+}$/Ne$^{1+}$&0.12&V)\\
e)& 11.37& S$^{14+}$& S$^{14+}$& Ne$^{2+}$/Ne$^{1+}$&0.06&V)\\
f)& 11.37& S$^{14+}$& S$^{14+}$& Ne$^{3+}$/Ne$^{1+}$&0.09&V)\\
g)& 0.15& He$^{2+}$& He$^{1+}$& Ne$^{1+}$/Ne$^{1+}$&0.38&III)\\
h)& 0.15& He$^{2+}$& He$^{1+}$& Ne$^{2+}$/Ne$^{1+}$&0.18&III)\\
i)& 0.15& He$^{2+}$& He$^{1+}$& Ne$^{3+}$/Ne$^{1+}$&0.21&III)\\
j)& 0.15& He$^{2+}$& He$^{0}$& Ne$^{2+}$/Ne$^{1+}$&0.61&III)\\
k)& 0.15& He$^{2+}$& He$^{0}$& Ne$^{3+}$/Ne$^{1+}$&0.27&III)\\
l)& 0.15& He$^{2+}$& He$^{1+}$& Ar$^{1+}$/Ar$^{1+}$&0.56&IV)\\
m)& 0.15& He$^{2+}$& He$^{1+}$& Ar$^{2+}$/Ar$^{1+}$&0.24&IV)\\
n)& 0.15& He$^{2+}$& He$^{1+}$& Ar$^{3+}$/Ar$^{1+}$&0.24&IV)\\
\hline
\end{tabular}
\caption[channels]{The studied collision systems and fragmentation channels are given for the used projectiles and target gases. In addition the ratio of ICD electrons to continuous electrons as described in Fig. \ref{ee} is given.}
\label{channels}
\end{center}
\end{table}

\subsection*{Assignment of ICD in the neon dimer}
The neon dimer is considered a model system for ICD. Numerous theoretical and experimental works \cite{santra2003,jahnke2004,scheit2004,santra2005,aoto2006,jahnke2007b,stoychev2008,demekhin2008,kreidi2008a,yamazaki2008,kreidi2008b,kreidi2009} have been performed. Experimentally ICD like mechanisms which lead to the fragmentation channels Ne$^{1+}$/Ne$^{1+}$ and Ne$^{2+}$/Ne$^{1+}$ have been identified after irradiation by synchrotron light. These are also found in nearly all collision systems studied here. In addition we find fragmentation into Ne$^{3+}$/Ne$^{1+}$, which was up to now not mentioned in the literature.

\subsubsection*{Ne$^{1+}$/Ne$^{1+}$}
The above described "model ICD mechanism" is the main contributor to the low energy electrons seen in Fig. \ref{ee}a - 2d and Fig. \ref{ee}g with the neon dimer fragmenting into Ne$^{1+}$/Ne$^{1+}$. It can thus be seen that ICD prominently appears in all relevant channels in ion/dimer collisions. These include pure ionization (a), projectile electron loss (b) and electron transfer (c), the latter is triggered by a He$^{1+}$ projectile. Qualitatively this behavior is unchanged for the single electron transfer process to the doubly charged projectile He$^{2+}$ given by (g). The only collision system studied in this work which does not exhibit ICD is the double electron transfer to the He$ ^{2+}$ projectiles with Ne$^{1+}$/Ne$^{1+}$ breakup for reasons of charge conservation. We note that the additional low energy electron contribution for fast (v = 21.2 a.u.), highly charged S$^{14+}$ projectiles (d) is clearly visible but substantially reduced compared to (a). 

\begin{figure*}[ht]
\centerline{\psfig{figure=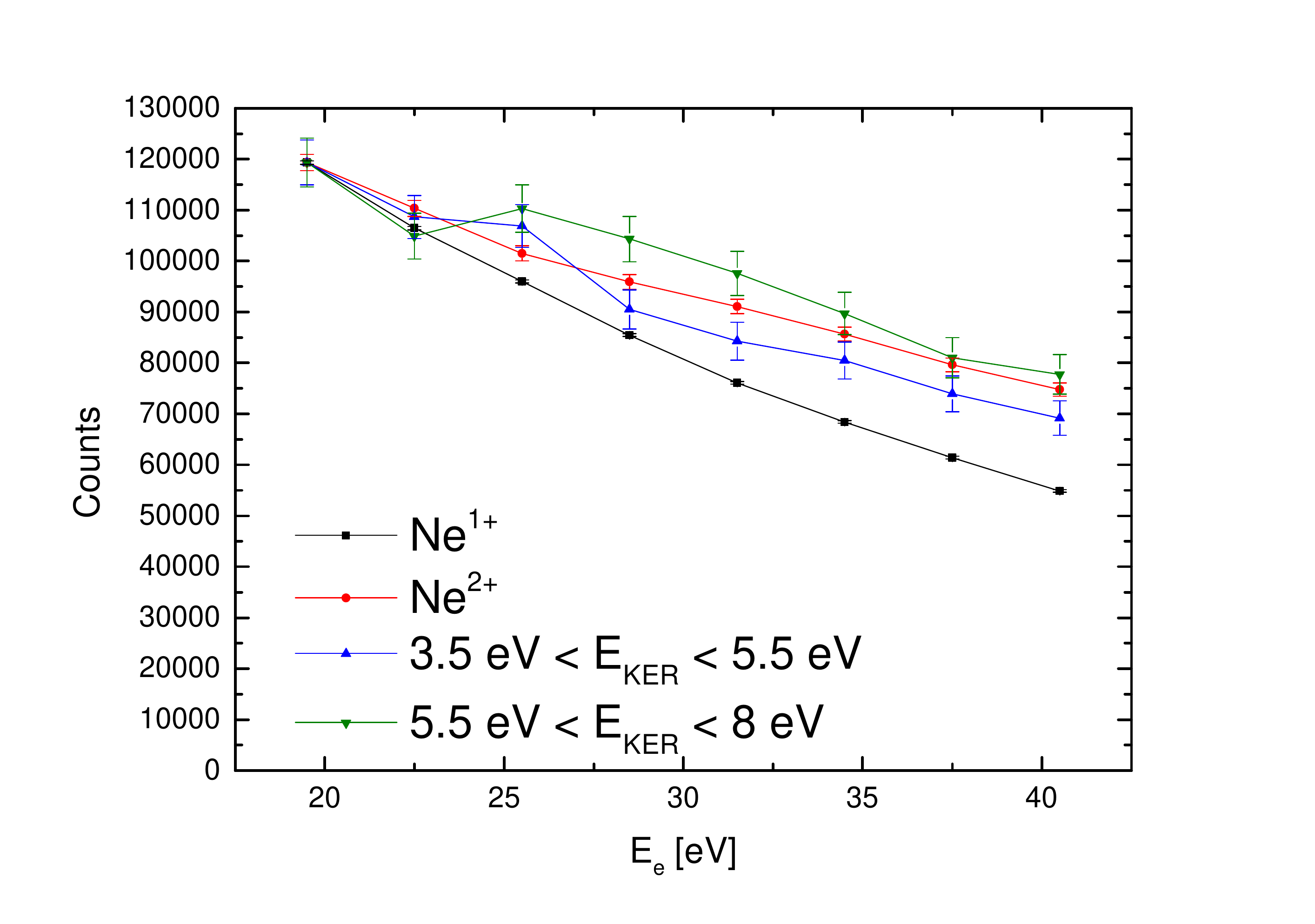,width=0.5\textwidth}}
\caption{(Color online) For channel (d) the projection of the electron energies E$_e$ for the two regions in the kinetic energy release 3.5 eV \textless\ E$_{KER}$ \textless\ 5.5 eV (blue up-pointing triangles) and 5.5 eV \textless\ E$_{KER}$ \textless\ 8 eV (green down-pointing triangles) are shown. For comparison the electron energy distributions for singly ionized neon atoms (black squares) and doubly ionized neon atoms (red circles) are depicted. All curves are scaled to the leftmost data point of the singly ionized monomers.}\label{comparison}
\end{figure*}

Fig. \ref{comparison} shows for channel (d) the projection of the electron energies for the two regions in the kinetic energy release 3.5 eV \textless\ E$_{KER}$ \textless\ 5.5 eV (blue up-pointing triangles) and 5.5 eV \textless\ E$_{KER}$ \textless\ 8 eV (green down-pointing triangles). For comparison the electron energy distributions for singly ionized neon atoms (black squares) and doubly ionized neon atoms (red circles) are depicted. All curves are scaled to the leftmost data point of the singly ionized monomers. 
It can be seen that the electron energy distribution for 5.5 eV \textless\ E$_{KER}$ \textless\ 8 eV, where the RCT process is initiated by the double ionization of one atomic center of the dimer, shows a similar decay behavior as the electron energy distribution of the doubly ionized monomers. 
The electron energy distribution for the range in the kinetic energy release 3.5 eV \textless\ E$_{KER}$ \textless\ 5.5 eV which is populated by single ionization processes followed by CE and ICD exhibits a significantly less steeper decay than the electrons that stem from the singly ionized monomers. This discrepancy might be caused by the dominant contribution of 2s-electrons which are emitted in the ICD process.

\subsubsection*{Ne$^{2+}$/Ne$^{1+}$}
\begin{figure*}[ht]
\centerline{\psfig{figure=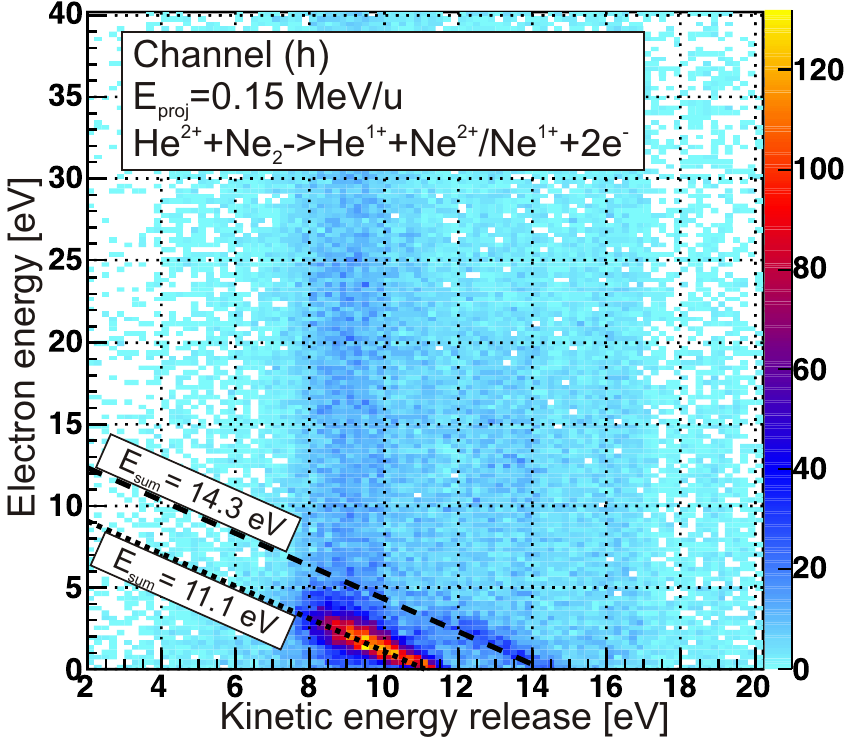,width=0.5\textwidth}}
\caption{(Color online) Correlation of electron and ion energies for channel (h). Horizontal axis: sum energy of both recoil ions (kinetic energy release), vertical axis: electron energy. Events from ICD are distributed along a diagonal line (see text).} \label{channelh}
\end{figure*}

The low energy electrons in the breakup channels given by Ne$^{2+}$/Ne$^{1+}$ ion pairs are attributed predominantly to two ICD mechanisms which were suggested by Santra et al. \cite{santra2003,santra2005} and Demekhin et al. \cite{demekhin2008} and observed by Kreidi et al. \cite{kreidi2008a,kreidi2008b} in relaxation processes from the Auger decay final state Ne$^{2+}$(2s$^{-1}$2p$^{-1}$)[$^{1}$P]/Ne[$^1$S] following 1s photoionization. In the first scenario, a 2p-electron from the doubly charged ion fills the 2s-hole where the excitation energy is transferred to the neighbor by a virtual photon. There a 2p-electron is removed leading to the dimer state Ne$^{2+}$(2p$^{-2}$)[$^{1}$D]/Ne$^{1+}$(2p$^{-}$)[$^{2}$P]. This process termed "direct ICD" exhibits a sum energy of E$_{sum}$ = E$_{KER}$ + E$_{e}$ = 11.1 eV and occurs at distances around the equilibrium bond length of the neutral neon dimer. In the second scenario, an electron of the neutral atom fills the 2s-hole of the doubly charged ion where simultaneously a second 2p-electron of the same site is ionized by the released excess energy. This "electron exchange" ICD, which ends in the final state Ne$^{2+}$(2p$^{-2}$)[$^{1}$P]/Ne$^{1+}$(2p$^{-}$)[$^{2}$P], has been reported to depend strongly on the spatial overlap of the wavefunctions \cite{jahnke2007b}. This is why it is opened only for small internuclear distances, hence leading to greater kinetic energy releases. For this relaxation pathway the sum energy is E$_{sum}$ = 14.3 eV. In Fig. \ref{channelh} (here shown for channel (h) but also visible in (e) and (j)) the sum energies of 11.1 eV for the direct ICD and 14.3 eV for the electron exchange ICD are indicated by a black dotted line and a black dashed line, respectively. Low energy electrons distributed along these lines clearly show IC decay mechanisms leading to Ne$^{2+}$/Ne$^{1+}$ after ion impact. 

\subsubsection*{Ne$^{3+}$/Ne$^{1+}$}
\begin{figure*}[ht]
\centerline{\psfig{figure=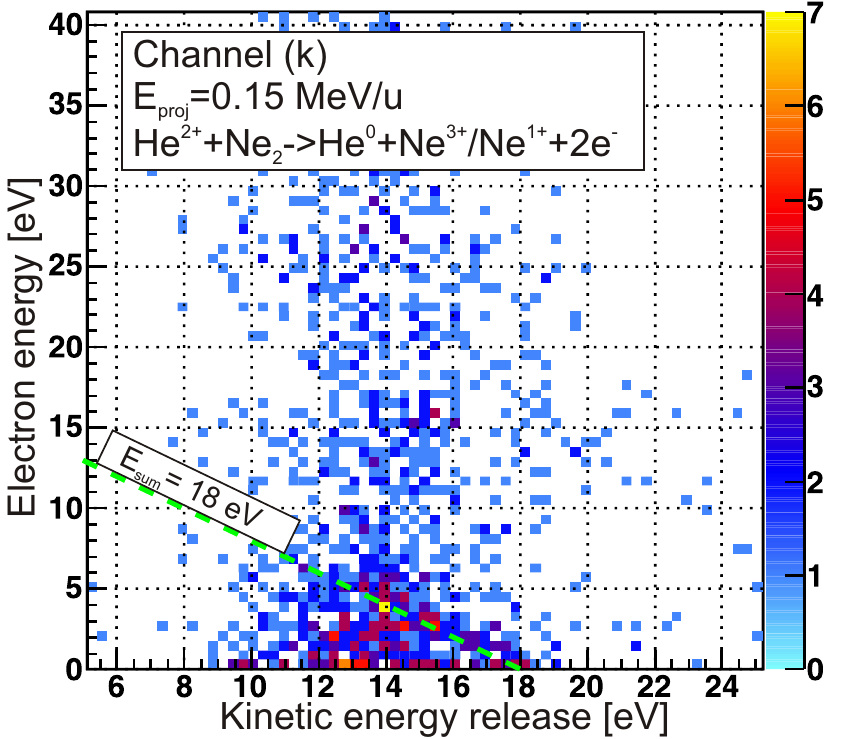,width=0.5\textwidth}}
\caption{(Color online) Correlation of electron and ion energies for channel (k). Horizontal axis: sum energy of both recoil ions (kinetic energy release), vertical axis: electron energy. Events from ICD are distributed along a diagonal line (see text).} \label{channelk}
\end{figure*}

\begin{figure*}[ht]
\centerline{\psfig{figure=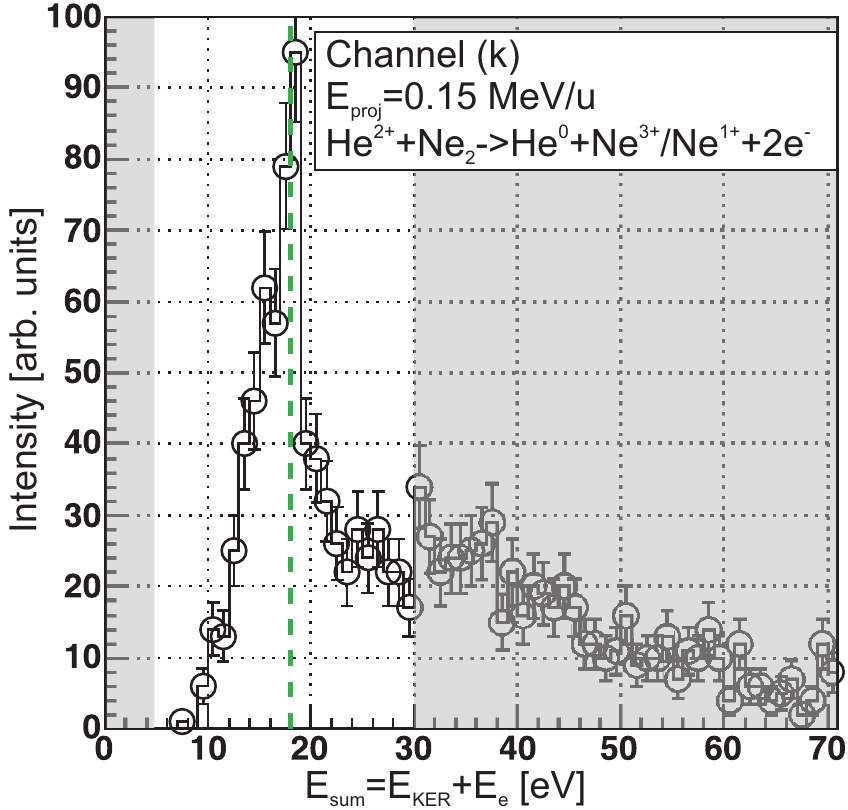,width=0.5\textwidth}}
\caption{(Color online) The sum energy (E$_{sum}$ = E$_{KER}$ + E$_{e}$) distribution is shown for channel (k) for 5 eV \textless\ E$_{KER}$ \textless\ 30 eV . A peak can be distinguished around E$_{sum}$ = 18 eV. The vertical green dashed line corresponds to the diagonal in Fig. \ref{channelk}.} \label{sum}
\end{figure*}

So far no theoretical prediction or experimental evidence was found in the literature which document ICD leading to the final ionic state Ne$^{3+}$/Ne$^{1+}$. In Fig. \ref{channelk} the E$_{KER}$ versus electron energy histogram is shown for channel (k). Apart from CE events centered around E$_{KER}$ $\approx$ 13 eV an accummulation of events along the sum energy E$_{sum}$\ $\approx$ 18 eV, indicated by the green dashed line, also suggest some type of ICD. In a one-dimensional representation of the sum energy in Fig. \ref{sum} a peak between 16 eV and 19 eV can be clearly seen.

\subsection*{Assignment of ICD in the argon dimer}
ICD was shown to occur also in argon dimers as shown in synchrotron radiation studies \cite{hergenhahn2011}. Depending on the initial state of excitation the final charge state of the fragmenting ion pair varies, whereby breakup channels into Ar$^{1+}$/Ar$^{1+}$ \cite{lablanquie2007,kimura2013}, Ar$^{2+}$/Ar$^{1+}$ \cite{morishita2006,ueda2007} and also Ar$^{3+}$/Ar$^{1+}$ \cite{sakai2011} were observed. In ion collision studies between slow, highly charged Ar$^{9+}$ projectiles and argon dimers \cite{matsumoto2010,matsumoto2011} also the breakup channels Ar$^{1+}$/Ar$^{1+}$, Ar$^{2+}$/Ar$^{1+}$, Ar$^{3+}$/Ar$^{1+}$ and Ar$^{2+}$/Ar$^{2+}$ were seen. In these studies ICD was suggested not to play a role due to electron capture mechanisms from the valence shell prohibiting the emission of an ICD electron. In our studies in fast collisions however we find reaction channels where at least one electron is emitted, thus opening ICD as a fast relaxation mechanism. The experimental resolution in the sum energy does however not suffice to observe any discrete diagonal structures for the argon dimers.

\subsubsection*{Ar$^{1+}$/Ar$^{1+}$}

\begin{figure*}[ht]
\centerline{\psfig{figure=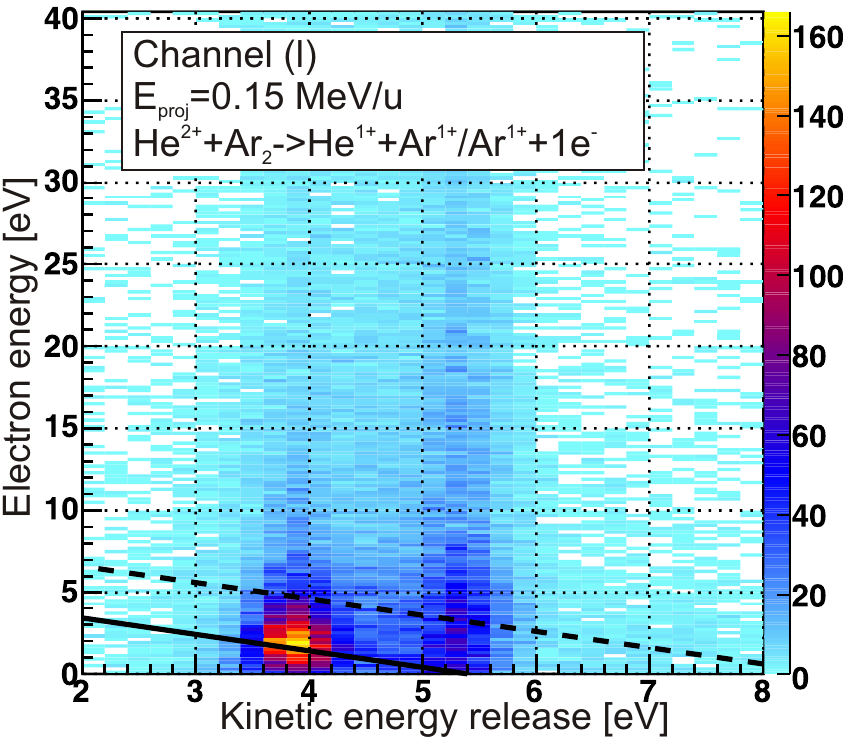,width=0.5\textwidth}}
\caption{(Color online) Correlation of electron and ion energies for channel (l). Horizontal axis: sum energy of both recoil ions (kinetic energy release), vertical axis: electron energy. Events from ICD are distributed along a diagonal line (see text).} \label{channell}
\end{figure*}

ICD in argon dimers with fragmentation into Ar$^{1+}$/Ar$^{1+}$ was reported by Lablanquie et al. \cite{lablanquie2007} by measuring the intensity of high energetic (from Coulomb explosion) Ar$^{+}$ ions after photon irradiation (E$_{\gamma}$ = 25 - 40 eV). Several satellite states of the singly ionized argon dimer Ar$^{+*}$-Ar were assumed to relax via ICD. Kimura et al. \cite{kimura2013} showed by a coincident ion-ion-electron measurement that a variety of states are also created by resonant Auger decay in argon dimers which then decay via ICD fragmenting to Ar$^{+}$(3p$^{-1}$)-Ar$^{+}$(3p$^{-1}$). The main contributions were assigned to the channels Ar$^{+}$[3p$^{-2}$($^{1}$D) 3d $^{2}$D]-Ar $\rightarrow$ Ar$^{+}$(3p$^{-1}$)-Ar$^{+}$(3p$^{-1}$) (E$_{sum}$ = 5.4 eV) and Ar$^{+}$[3p$^{-2}$($^{1}$D) 4d $^{2}$D]-Ar $\rightarrow$ Ar$^{+}$(3p$^{-1}$)-Ar$^{+}$(3p$^{-1}$) (E$_{sum}$ = 8.6 eV) indicated by the solid and dashed lines in Fig. \ref{channell}, respectively. Although intense "islands" can be clearly distinguished along these two lines the broadness of the structures indicates that many more satellite states of the dimer are populated which cannot be resolved.

\subsubsection*{Ar$^{2+}$/Ar$^{1+}$}

\begin{figure*}[ht]
\centerline{\psfig{figure=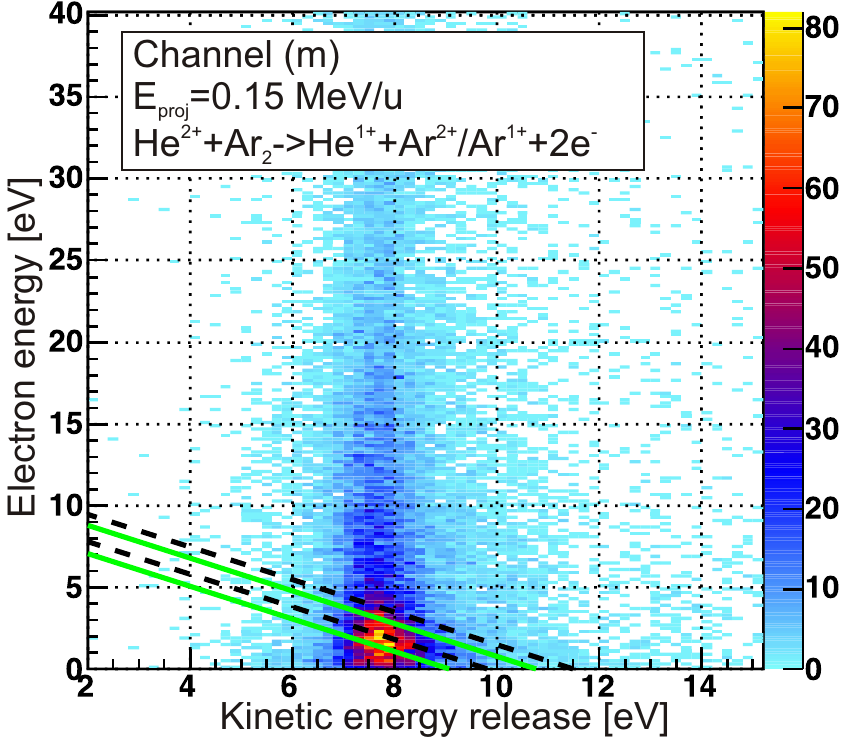,width=0.5\textwidth}}
\caption{(Color online) Correlation of electron and ion energies for channel (m). Horizontal axis: sum energy of both recoil ions (kinetic energy release), vertical axis: electron energy. Events from ICD are distributed along a diagonal line (see text).} \label{channelm}
\end{figure*}

ICD with breakups into Ar$^{2+}$/Ar$^{1+}$ ions are reported to originate mainly from Ar$^{2+}$(3p$^{3}$)-Ar  states which then decay into Ar$^{2+}$(3p$^{-2}$)-Ar$^{1+}$(3p$^{-1}$) states \cite{morishita2006,ueda2007}. In particular the events from the relaxation paths denoted by Ar$^{2+}$(3p$^{-3}$3d)[$^{1}$P]-Ar $\rightarrow$ Ar$^{2+}$(3p$^{-2}$)[$^{3}$P]-Ar$^{1+}$(3p$^{-1}$)[$^{2}$P] (E$_{sum}$ = 11.5 eV) and Ar$^{2+}$(3p$^{-3}$3d)[$^{1}$P]-Ar $\rightarrow$ Ar$^{2+}$(3p$^{-2}$)[$^{1}D$]-Ar$^{1+}$(3p$^{-1}$)[$^{2}$P] (E$_{sum}$ = 9.8 eV) seem to significantly produce the low energy electrons. Their sum energies are marked in Fig. \ref{channelm} by black dashed lines.  Also the events from Ar$^{2+}$(3p$^{-3}$3d)[$^{3}$P]-Ar leading to the same final states are energetically located only 0.73 eV below the other channels in the sum energy (green solid lines in Fig. \ref{channelm}). Additional weak contributions from among others Ar$^{2+}$(3s$^{-2}$)[$^{1}$S]-Ar and Ar$^{2+}$(3p$^{-3}$$^{2}$D 4d$^{3}$P)-Ar fragmenting to the states Ar$^{2+}$(3p$^{-2}$)[$^{3}$P]-Ar$^{1+}$(3p$^{-1}$)[$^{2}$P], Ar$^{2+}$(3p$^{-2}$)[$^{1}D$]-Ar$^{1+}$(3p$^{-1}$)[$^{2}$P] and Ar$^{2+}$(3p$^{-2}$)[$^{1}S$]-Ar$^{1+}$(3p$^{-1}$)[$^{2}$P] via ICD are assumed to add to the yield of electrons up to energies around ~10 eV \cite{ueda2007}. 
In Fig. \ref{channelm} the data is located between 6 eV \textless\ E$_{KER}$ \textless\ 10 eV. Apart from the decay via direct Coulomb explosion (CE expected to occur around 7.6 eV) we attribute the accumulation of electrons at energies below roughly 10 eV to ICD.

\subsubsection*{Ar$^{3+}$/Ar$^{1+}$}
In 2011 Sakai et al. \cite{sakai2011} showed that ICD occurs from triply charged states of the argon dimer Ar$^{3+}$-Ar resulting in Ar$^{3+}$-Ar$^{1+}$ states. They observed peaks of ~11 eV in the E$_{KER}$ (corresponding to the bond length of the ground state argon dimer r$_{0,Ar2}$ = 3.76 \AA\ \cite{stoychev2008b,patkowski2005} within the reflection approximation)  and at about ~ 1 eV in the electron energy which were attributed to exchange ICD from Ar$^{3+}$(3s3p$^{4}$)[$^{2}$P]-Ar to Ar$^{3+}$(3p$^{3}$)[$^{4}$S]-Ar$^{1+}$(3p$^{5}$)[$^{2}$P]. Also direct ICD from various satellite states were reported to be energetically allowed. Similar to the Ar$^{2+}$/Ar$^{1+}$ case the peak in Fig. \ref{ee}n is attributed to ICD events.

\section{Conclusions}
In conclusion we report of an increased low energy electron yield in a broad range of ion/Ne$_{2}$ and ion/Ar$_{2}$  collisions. The projectiles differ in charge state and velocity as well as the change of the charge state after the reaction which arises from pure ionization, projectile electron loss, single and double electron capture. This surplus of electrons can be unambiguously assigned to ICD in the case of Ne$^{1+}$/Ne$^{1+}$ and Ne$^{2+}$/Ne$^{1+}$. For all other channels ICD was also shown to be responsible for the enhanced emission of low energy electrons. Except for the Ne$^{1+}$/Ne$^{1+}$ fragmentation after ionization \cite{kim2011} ICD was observed for the first time in all remaining ion/dimer collisions. Our comprehensive study suggests that in fast ion collisions with loosely bound matter ICD is omnipresent. It is a major contributor to the creation of low energy electrons.

\begin{acknowledgments}
This work has been supported by DFG and by the European Community FP7 - Capacities- Integrated Infrastructure Initiative- contract ENSAR n$^\circ$ 262010.  We gratefully acknowledge the GANIL and CIMAP staff for their exceptional help during the experiment in Caen and  K. Stiebing, P. Ziel, W. Dilfer, and M. Dworak for the excellent preparation of the ion beam in Frankfurt.

\end{acknowledgments}

\newpage

\end{document}